\def\year{2020}\relax
\documentclass[letterpaper]{article} % DO NOT CHANGE THIS
\usepackage{aaai20}  % DO NOT CHANGE THIS
\usepackage{times}  % DO NOT CHANGE THIS
\usepackage{helvet} % DO NOT CHANGE THIS
\usepackage{courier}  % DO NOT CHANGE THIS
\usepackage[hyphens]{url}  % DO NOT CHANGE THIS
\usepackage{graphicx} % DO NOT CHANGE THIS
\usepackage{graphicx}  % DO NOT CHANGE THIS
\usepackage{enumitem} 
\usepackage{multirow} 
\usepackage{hhline} 
\usepackage{booktabs} 
\usepackage{subcaption} 
\usepackage{stackengine}

\usepackage{soul}
\usepackage{color}
\usepackage{todonotes}
\usepackage{amsmath,amssymb}

\colorlet{ColorforSina}{green!5!orange!95!}

\title{Quantitative Evaluation of Machine Learning Explanations: \\ A Human-Grounded Benchmark}

\author{Sina Mohseni,\textsuperscript{\rm 1}
Jeremy E. Block,\textsuperscript{\rm 2}
Eric D. Ragan, \textsuperscript{\rm 2} \\
\textsuperscript{\rm 1} Texas A\&M University, College Station, Texas \\
\textsuperscript{\rm 2} University of Florida, Gainesville, Florida \\
\{sina.mohseni,\}@tamu.edu, \{j.block,eragan\}@ufl.edu
}

\begin{document}

\maketitle

% --- Proofreading --- 
%  - [done] update colors in the last chart
%  - [done] check the consistancy between past and present tense  
% -  [nope] citing the workshop paper in the full paper? 
% -  [done] size of the supplementary material
%  - [done] Remove the section numbering style
% - \jeremy{add/adjust/ clarify the working definition for "human evaluation rating"}

% ----- Results -------
% - [done] regression slopes 
% - [done]  update LIME images 
    % - \jeremy{are these a selection of the best of each explainer? I feel like it might be good to have one or two that were rated poorly by users? maybe one where LIME was better and one where GRAD-CAM is better?}

% ----- References/Def -------
% - [nope]  Intersection over Union (IoU) and mean Average precision (mAP)
% - baseline vs. benchmark
% - FN vs FN errors
% - participant vs annotator

% ----- Benchmark -------
% - how did authors performed the seg mask
% - How many participant where required total?

% ----- Figures -------
% - adding weights and biases to figures caption 
% - change the horse figure. 
% - change the cat and dog figures

% - Update results: 
    % - add lime results 
    % - compare first half and second half of the user rating result
% - [done] Update figures 
% - [done] write the discussion 
% - [done] Image annotation: 3 done - 17 remaining
    % - [done] run+clean up
% - [done] Image rating - LIME -
% - [done] Text annotation: get the IMDB - 
     %  - [done] (sat) running 
     %  - (sun) clean up...

\begin{abstract}

Research in interpretable machine learning proposes different computational and human subject approaches to evaluate model saliency explanations. 
These approaches measure different qualities of explanations to achieve diverse goals in designing interpretable machine learning systems. 
In this paper, we propose a human attention benchmark for image and text domains using multi-layer human attention masks aggregated from multiple human annotators.  
We then present an evaluation study to evaluate model saliency explanations obtained using Grad-cam and LIME techniques. 
We demonstrate our benchmark's utility for quantitative evaluation of model explanations by comparing it with human subjective ratings and ground-truth single-layer segmentation masks evaluations. 
Our study results show that our threshold agnostic evaluation method with the human attention baseline is more effective than single-layer object segmentation masks to ground truth. 
Our experiments also reveal user biases in the subjective rating of model saliency explanations. 
% that users more severely punish subjective ratings of the false positive elements in an explanation then other error types. 
% with single-layer feature mask and human review/judgment of explanations and discuss the difference.. . 

\end{abstract}

\section{Introduction} 

%-- what is XAI --
\noindent With the recent and continuing advancements in robust deep neural networks (DNN), the prominence of machine learning techniques for automated decision-making is growing.
In such cases, human experts, operators, and decision-makers can also take advantage of advanced machine learning techniques to account for latent features and assist in taking real-world actions. 
However, because of the disparity between the sense-making process in humans and the computational feature learning of machine learning models, people require model transparency to be able to understand and trust machine learning models. 
Thus, for more effective human-AI collaboration, advancements in model explainability are needed to support human understanding. 
This is the primary goal of recent interdisciplinary research thrusts in \textit{Explainable Artificial Intelligence} (XAI).
While a multi-faceted topic, the ultimate goal is for people to understand machine models, and it is therefore essential to involve human feedback and reasoning as a requisite component for design and evaluation of XAI systems~\cite{mohseni2019survey}.

% why is evaluation of explanation important.
Research on interpretable algorithms has recently proposed various techniques to design inherently interpretable models~\cite{wu2018beyond} and generate explanations for black-box models~\cite{selvaraju2017grad}. 
Interpretability techniques enable human review of model reasoning and learning representations for their correctness in accordance to design goals, law and regulations, and safety requirements. 
Such evaluations could potentially prevent adverse outcomes of AI-based systems---such as unfair and discriminatory decision-making when performing real-world tasks. 
However, with the complexity of interpretability techniques and human cognitive biases, the question remains: how should we choose effective and efficient methods for the evaluation of machine learning explanations?

% what are methods to evaluate 
Different approaches have been proposed for evaluating interpretable models and XAI systems at different stages of system design~\cite{doshi2017towards}. 
In machine learning research, various computational methods are used to measure the fidelity of interpretability techniques with respect to the underlying black-box model~\cite{adebayo2018sanity,hooker2019benchmark}. 
On the other hand, in the field of human-computer interaction, human-grounded evaluation approaches measure human factors such as user satisfaction, mental model, and trust in XAI systems designed for different tasks.  % ~\cite{mohseni2019survey}
% transition sentence needed
However, there are fundamental differences between these evaluation approaches. 
Computational methods set a precedent to objectively evaluate the model against a baseline ground truth, yet they lack the ability to quantify human interpretations. 
On the other hand, while more descriptive in nature, human subject studies tend to be more costly, imprecise, and subjective to the task.
Another major difference between these evaluation methods is that once the human user is exposed to the evaluation study setup, she can not unlearn the experience for another round of evaluation. 
These differences raise the need to study the trade-off between objective ground-truth evaluation and subjective human-judgment of explanations.

% goal and contributions. 
In this paper, we propose a human-attention baseline to quantitatively evaluate model saliency explanations. 
Our publicly available~\footnote{https://github.com/SinaMohseni/ML-Interpretability-Evaluation-Benchmark} % Anonymous Link (a subset of the benchmark is submitted as supplementary material) 
human-grounded benchmark enables fast, replicable, and objective execution of evaluation experiments for saliency explanations. 
To foster the interest of the machine learning community, we demonstrate our benchmark's utility for quantitative evaluation of model explanations and compare it with the single-layer feature mask ground truth and human judgment rating evaluations.
Our study results reveal the efficiency of threshold-agnostic evaluation with a human-attention baseline as compared to previous methods with binary ground truth masks and labels. 
Our experiments also reveal user biases between different model error types in the subjective rating of explanations.

\section{Background}

The evaluation of model explanations and interpretability techniques can be categorized in different ways~\cite{doshi2017towards,mohseni2019survey}.
For instance, previous works have examined the fidelity of interpretability techniques to the black-box model~\cite{hooker2019benchmark,adebayo2018sanity}, evaluated correctness of model explanations with ground-truth~\cite{du2018towards}, as well as the usefulness of explanations in different tasks and domains~\cite{kocielnik2019will}.
In this paper, we focus on the trustworthiness of explanations with the assumption of having a high-fidelity ad-hoc explainer.
In this section, we review two evaluation approaches, \textit{human judgment evaluation} and \textit{ground-truth evaluation}, for the trustworthiness of machine learning explanations and assess their advantages and limitations.

\subsection{Human Judgment Evaluation} 
% \subsection{Subjective Evaluation} 

A common approach for evaluating machine learning explanations is the direct review of model explanations with end-users for their subjective feedback. 
Multiple papers have reported measurements of users' understanding of explanations as a proxy for usefulness and interpretability of explanations~\cite{lage2019human,poursabzi2018manipulating}. 
Others have measured user-reported trust as a proxy for explanation goodness.
For example, Nourani et al.~\shortcite{nourani2019effects} and Papenmeier et al.~\shortcite{papenmeier2019model} studied the effects of explanation meaningfulness and ad-hoc explainer fidelity on user reliance. 
Both studies show that model accuracy and explanation fidelity impact users' trust in the model and conclude that providing nonsensical explanations (i.e., those that do not align with users' expectations) may harm users' reported trust and observed reliance on the system. 
With a crowdsourced evaluation approach, Schmidt and Biessmann~\shortcite{schmidt2019quantifying} present quantitative measures for system interpretability and human trust. 
They propose that analyzing user interaction time can serve as a proxy for users' understanding of the explanation and level of trust. 
They recommend that model explanations need to enhance the information transfer rate to users, help users establish an intuitive understanding of system performance and perform well independent from the user task. 

Taking a different perspective, Schneider et al.~\shortcite{schneider2020deceptive} inspected the effects of deceptive model explanations in a user study. 
Their findings indicate that explanations that are unfaithful to the black-box model can fool users in accepting wrong predictions. 
Following a similar goal, Lakkaragu et al.~\shortcite{lakkaraju2019fool} present an approach to generate misleading explanations and a case study with law and criminal justice domain experts. 
Their study results found that misleading explanations were able to significantly increase users' trust. 
Conclusively, various research efforts have shown the limitations of human judgment for robust evaluation of machine learning explanations.

Different projects have run user studies to evaluate the human understanding of saliency map explanations from DNNs. 
For example, Alqaraawi et al.~\shortcite{alqaraawi2020evaluating} showed that instance explanations carry new information to users, but model behavior remained largely unpredictable for participants. 
In other work, Zhang et al.~\shortcite{Zhang2019dissonance} compared saliency explanations from multiple networks with human explanations of objects in images. 
They performed a large crowdsourced study to directly compare machine learning and human explanations and human feedback on model explanations. 
Their results indicate that the features learned by some DNN models are more similar to human intuition. 
However, it is not clear from their study whether the model generalizability or the choice of interpretability technique was more effective on user satisfaction of explanations. 
To address the limitations in human judgment evaluation studies, Lertvittayakumjorn and Toni~\shortcite{lertvittayakumjorn2019human} defined a set of objective evaluation tasks for quantitative evaluation of
model explanations with respect to different explanatory purposes.
They used three human-grounded tasks to evaluate local explanation methods for their ability to reveal model behavior, justify model predictions, and help users investigate uncertain predictions.
The review of previous research indicates that the dissonance between machine learning models' goal to~\textit{learn discriminant features} and human expectation of \textit{logical and common sense explanations} undermines the correctness and completeness of human judgment evaluation methods.

% \eric{make explicit note with light mention of CHI workshop paper}

% To address this gap, Holzinger et al.~\cite{holzinger2020measuring} proposed the system causability scale for measuring the quality of explanations. 
% They provide likert scale survey to determine the suitability of an explanation type for the intended purpose. 

% while Adebayo et al. show that saliency maps are no better then an edge detection filter~\shortcite{adebayo2018sanity}. 

\subsection{Evaluation with Ground Truth} 
% \subsection{Human-Grounded Evaluation of XAI}
% \subsection{Direct Evaluation of XAI}

An objective way to quantify the correctness of model explanation is to examine it against a ground truth baseline. 
Ground truth is often defined by human annotation of representative features (i.e., feature masks) and provide a baseline for quantitative evaluation of explanations quality. 
Examples include annotations of the object's ``segmentation mask'' in natural datasets, e.g., ~\cite{Everingham15}, and synthesized datasets, e.g.,~\cite{osman2020towards}, that represent specific features associated with the target class.
Different similarity metrics, such as Intersection over Union (IoU) (also called Jaccard index) and mean Average Precision (mAP), are used to quantify the quality of model saliency explanations or bounding boxes compared to the ground truth. 
For instance, Li et al.~\shortcite{li2018tell} used IoU, between the model saliency map from a Convolutional Neural Network (CNN) and the ground truth segmentation mask from the validation set, to measure their quality as a weakly-supervised semantic segmentation task. 
In another work, Du et al.~\shortcite{du2018towards} calculate the mAP between the bounding boxes of an objects' saliency mask and the ground truth bounding boxes to evaluate their interpretability technique as an object localization task.
Similarly, in the text domain, direct comparison of model attention explanations with human annotated sentences, e.g., evidence supporting the target label~\cite{zaidan2007using}, provides an explanation quality score~\cite{lei2016rationalizing}.
However, the relationship between the evaluation of machine learning explanations and the auxiliary tasks, such as binary object localization and semantic segmentation, is not clear yet.
Our research highlights the aspects of \textit{human feedback} that are missing in the \textit{human annotation} baselines but would complement the evaluation of machine learning explanations.

In a review of limitations in threshold-based evaluations for model saliency map, Choe et al.~\shortcite{choe2020evaluating} present an evaluation protocol to include a hyperparameter search for the $\tau$ threshold for generating objects' ``binary mask'' from the saliency score map. 
However, unlike our proposed evaluation protocol, they do not consider the pixel-wise evaluation of saliency score maps in the first place. 
Aside from object segmentation mask baselines that annotate entire features associated with the target class, perhaps closest work to our human attention benchmark is Das et al.'s~\shortcite{das2017human} VQA-HAT baseline for evaluating saliency maps in visual question and answering models. 
They test multiple game-inspired, attention annotation methods to ask participants to sharpen regions of a blurred image to answer a question. 
The resulting baseline is a human attention map that enables object identification but does not indicate whether the necessary or sufficient features are annotated by individual participants.
In comparison to VQA-HAT, our benchmark assembles annotations from multiple unique participants to generate accurate and generalizable human-attention maps. 
In this paper, we present a series of evaluation experiments and argue that our proposed multi-layer human-attention baseline is able to evaluate the completeness (i.e., the existence of false-negative explanation errors) and correctness (i.e., the existence of false-positive explanation errors) of model saliency maps.

\begin{figure*}[t!]
\centering
    \begin{subfigure}[b]{0.25\columnwidth}
        \includegraphics[width=0.90\columnwidth]{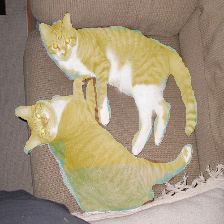}
        % \caption{}
        \\
    %   \hfill
    \end{subfigure}
     \begin{subfigure}[b]{0.25\columnwidth}
        \includegraphics[width=0.90\columnwidth]{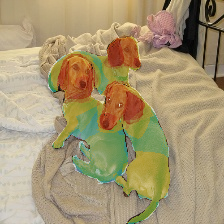}
        % \caption{}
        \\
    \end{subfigure}
    %   \hfill
      \begin{subfigure}[b]{0.25\columnwidth}
        \includegraphics[width=0.90\columnwidth]{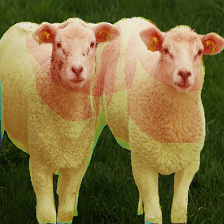}
        % \caption{}
        \\
    \end{subfigure}
    %   \hfill
    \begin{subfigure}[b]{0.25\columnwidth}
        \includegraphics[width=0.90\columnwidth]{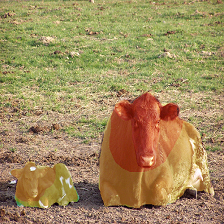}
        % \caption{}
        \\
    \end{subfigure}
    \begin{subfigure}[b]{0.25\columnwidth}
        \includegraphics[width=0.90\columnwidth]{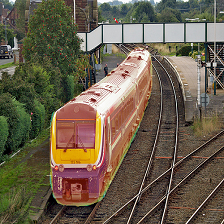}
        % \caption{}
        \\
    \end{subfigure}
    %   \hfill
     \begin{subfigure}[b]{0.25\columnwidth}
        \includegraphics[width=0.90\columnwidth]{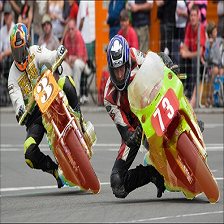}
        % \caption{}
        \\
    \end{subfigure}
    %   \hfill
      \begin{subfigure}[b]{0.25\columnwidth}
        \includegraphics[width=0.90\columnwidth]{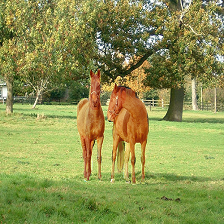}
        % \caption{}
        \\
    \end{subfigure}
    %   \hfill
    \begin{subfigure}[b]{0.25\columnwidth}
        \includegraphics[width=0.90\columnwidth]{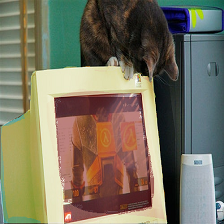}
        % \caption{}
        \\
    \end{subfigure}
    \\
    \begin{subfigure}[b]{0.25\columnwidth}
        \includegraphics[width=0.90\columnwidth]{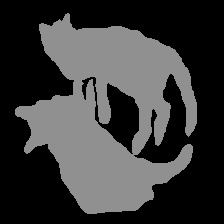}
        % \caption{}
        \\
    %   \hfill
    \end{subfigure}
     \begin{subfigure}[b]{0.25\columnwidth}
        \includegraphics[width=0.90\columnwidth]{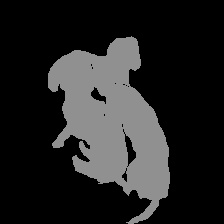}
        % \caption{}
        \\
    \end{subfigure}
    %   \hfill
      \begin{subfigure}[b]{0.25\columnwidth}
        \includegraphics[width=0.90\columnwidth]{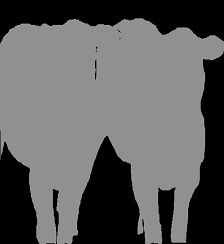}
        % \caption{}
        \\
    \end{subfigure}
    %   \hfill
    \begin{subfigure}[b]{0.25\columnwidth}
        \includegraphics[width=0.90\columnwidth]{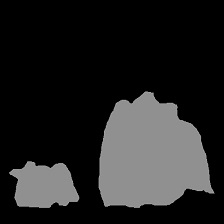}
        % \caption{}
        \\
    \end{subfigure}
    \begin{subfigure}[b]{0.25\columnwidth}
        \includegraphics[width=0.90\columnwidth]{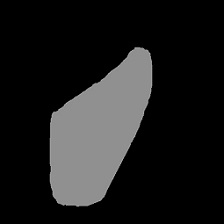}
        % \caption{}
        \\
    \end{subfigure}
    %   \hfill
     \begin{subfigure}[b]{0.25\columnwidth}
        \includegraphics[width=0.90\columnwidth]{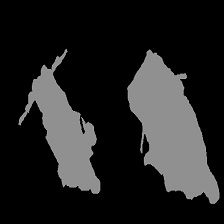}
        % \caption{}
        \\
    \end{subfigure}
    %   \hfill
      \begin{subfigure}[b]{0.25\columnwidth}
        \includegraphics[width=0.90\columnwidth]{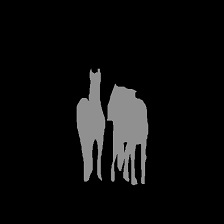}
        % \caption{}
        \\
    \end{subfigure}
    %   \hfill
    \begin{subfigure}[b]{0.25\columnwidth}
        \includegraphics[width=0.90\columnwidth]{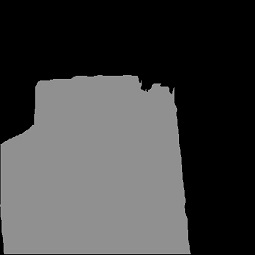}
        % \caption{}
        \\
    \end{subfigure}
    \\
    \begin{subfigure}[b]{0.25\columnwidth}
        \includegraphics[width=0.90\columnwidth]{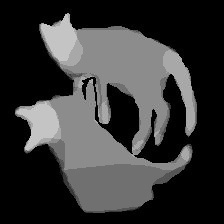}
        \caption{Cat}
        % \\ \centering Cat
    \end{subfigure}
    %   \hfill
     \begin{subfigure}[b]{0.25\columnwidth}
        \includegraphics[width=0.90\columnwidth]{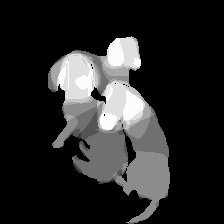}
        \caption{Dog}
        % \\ 
    \end{subfigure}
    %   \hfill
      \begin{subfigure}[b]{0.25\columnwidth}
        \includegraphics[width=0.90\columnwidth]{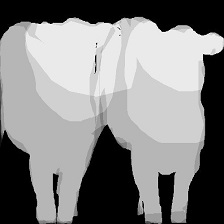}
        \caption{Sheep}
    \end{subfigure}
    %   \hfill
    \begin{subfigure}[b]{0.25\columnwidth}
        \includegraphics[width=0.90\columnwidth]{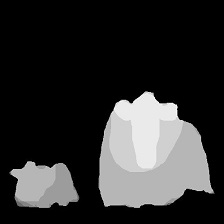}
        \caption{Cow}
    \end{subfigure}
    \begin{subfigure}[b]{0.25\columnwidth}
        \includegraphics[width=0.90\columnwidth]{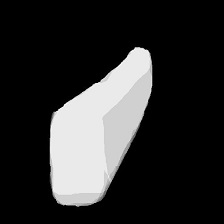}
        \caption{Train}
    \end{subfigure}
    %   \hfill
     \begin{subfigure}[b]{0.25\columnwidth}
        \includegraphics[width=0.90\columnwidth]{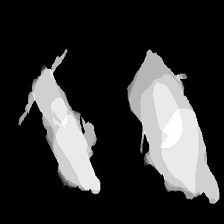}
        \caption{Motorbike}
    \end{subfigure}
    %   \hfill
      \begin{subfigure}[b]{0.25\columnwidth}
        \includegraphics[width=0.90\columnwidth]{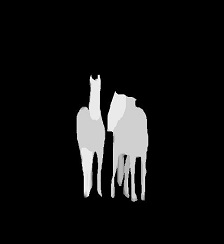}
        \caption{Horse}
    \end{subfigure}
    %   \hfill
    \begin{subfigure}[b]{0.25\columnwidth}
        \includegraphics[width=0.90\columnwidth]{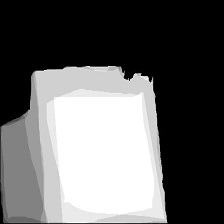}
        \caption{TV/Monitor}
    \end{subfigure}
  \caption{Examples of human annotations of salient features on images with the target class in the caption. \textbf{(Top)} Input images with human-attention mask heatmap overlay.  \textbf{(Middle)} Single-layer object's segmentation mask for the target class. \textbf{(Bottom)} Resulting multi-layer human attention mask. Each image is annotated by 10 unique participants.
  }
  \label{fig:main-figure}
\end{figure*}

\section{Human-Attention Benchmark}

We captured the human annotation of salient features in order to create a human-grounded benchmark to evaluate model explanations. 
Participants were prompted to select relevant regions in images and phrases in text documents that they felt most representative of the target subject or topic, respectively. 
Figure~\ref{fig:main-figure} show examples from the resulting multi-layer ground truth from aggregating annotation from multiple unique annotators for each image. 
In comparison to the single-layer object's segmentation map, the human-attention benchmark allows for a higher level of granularity in the evaluation of saliency maps and reflects human attention to features.
Also, compared to human judgment rating evaluations, the human attention benchmark enables reproducible and cost-efficient evaluation. 
The following reviews the details of benchmark specification, annotation procedure, and data processing.

% \usepackage{booktabs}
% \usepackage{graphicx}
\begin{table}[]
\centering
\caption{Details of the evaluation benchmark for human attention masks in different datasets. }
\label{tab:main-table}
\resizebox{\linewidth}{!}{%
\begin{tabular}{@{}ccccc@{}}
\toprule
Domain & \multicolumn{2}{c}{Image} & \multicolumn{2}{c}{Text} \\ \midrule
Dataset & \begin{tabular}[c]{@{}c@{}}PASCAL VOC \\ 2012\end{tabular} & \begin{tabular}[c]{@{}c@{}}ILSVRC \\ 2014\end{tabular} & 20 Newsgroup & \begin{tabular}[c]{@{}c@{}}IMDB 50K\end{tabular} \\ \midrule
Number of classes & 20 & 20 & 2 & 2 \\ \midrule
Samples per class & 50 & 5 & 100 & 100 \\ \midrule
\begin{tabular}[c]{@{}c@{}}Total annotation\\  sample size\end{tabular} & 1000 & 100 & 200 & 200 \\ \bottomrule
\end{tabular}%
}
\end{table}

\subsection{Benchmark Specifications}

The benchmark presents multi-layer masks representing what features humans expect to be the most important representations of a particular class. 
For each sample, we collect annotations from 10 unique annotators from Amazon Mechanical Turk platform that were instructed to select areas (in images) or words (in documents) that they deem most relevant to the target class. 
The multi-layer mask generated by aggregating annotations for each individual sample provides more granular representation of attributed features compared to the single-layer mask.  
Note that our method---collecting multiple user annotations for human-attention masks---balances the trade-off between objective annotation of precise feature-masks (i.e., segmentation mask) and subjective human judgment of the representative features. 
Also, it is important to mention that this human-attention baseline evaluates the explanations' correctness or trustworthiness of saliency explanations and does not intend to measure the fidelity of ad-hoc interpretability techniques to the black box models.

The development of this benchmark consists of a validation subset from \textit{ImageNet}~\cite{deng2009imagenet} and \textit{PASCAL VOC2012}~\cite{Everingham15} image datasets and \textit{20 Newsgroup}~\cite{Lang95newwgroup} and \textit{IMDB}~\cite{pang2004sentimental} text datasets. 
Table~\ref{tab:main-table} presents details for the number of classes and annotated samples from the four datasets in our explanation evaluation benchmark. 
For the PASCAL VOC dataset, 50 randomly selected samples from all 20 classes are annotated including Vehicles (airplane, bicycle, boat, boat, bus, car, motorbike, train), Households (bottle, chair, dining table, potted plant, sofa, TV/monitor), and Animals (bird, cat, cow, dog, horse, sheep) and other (person).
To create a validation set from the ImageNet dataset, we randomly selected five images from 20 classes including living things (man, woman, cat, dog, bird, ant, elephant, shark, zebra, flower, tree), indoor objects (chair, computer, ball, book, phone), outdoor objects (car, ship, airplane, house). 
The set includes images covering broad considerations such as multi-object and complex scenes, co-occurrence of target object, target object in different scales, and lighting conditions.

For the text domain datasets, 100 randomly selected movie reviews from each positive and negative classes of \textit{IMDB} dataset are selected. 
Similarly, 100 randomly selected text documents (with the headers removed from samples) from the \textit{20 Newsgroup} dataset are selected from two categories of medical (\textit{sci.med}) and electronic (\textit{sci.elect}).

\subsection{Annotation Interface and Procedure}

In order to generate multi-layer human-attention explanations, we ask annotators to provide their interpretations of the salient features that are most meaningful for the specific class from the data set. 
Each sample is annotated with 10 unique annotators recruited from Amazon Mechanical Turk (AMT). 
Recruitment advertisement for Human Intelligence Task (HIT) required participants to have at least 1000 previously approved HITs in AMT platform with the HIT approval rate of above 95\%.   % live in the US and have seen our advertisement.
Recruited participants were walked through a training slideshow of the task instructions and interface controls at the beginning of their HIT. 
As a control, each training slide was displayed on screen for two seconds before participants were able to continue to the next slide. 
Afterward, they were asked to agree to the IRB approved information sheet for data collection, and continued to a set of 12 images or documents for annotation. % this is down to 12
Participants were paid \$0.40 for the image and text annotation HITs to reach an average hourly pay rate of \$10 an hour.

We designed two fundamentally similar human annotation interfaces to capture human feedback for all image and text datasets.   
Annotators were using an interface with basic annotation tools in which each document or image was presented individually.    
Each annotation HIT started with the same two samples to serve as attention check and help the annotator to get adjusted with the interface and task.    
These are then followed by 10 samples from the main validation set.  
Task instructions prompted participants to select relevant regions in images that they felt most representative to the target object that could be entire or parts of it but generally not the background scenery. 
For image annotations, the annotators were specifically asked to use their mouse to lasso ``salient area(s) that explain target \emph{``object''} in the image''. % a series of images.  
Similarly, for text annotations, participants were prompted to select relevant words in text documents that they felt most representative of the target topic or class. 
For example, for the movie review IMDB dataset, the annotators were explicitly asked to ``select words and phrases which explain the positive or negative sentiment of the movie review''.

\subsection{Data Processing and Storage}

In order to generate multi-layer feature masks from multiple user annotations, we run a union operation on all individual annotation that displays what areas are most frequently selected by the annotators. 
Figure~\ref{fig:main-figure} presents examples of resulting human-attention masks from different images. 
Although specified in annotation task instructions, we also applied the exact segmentation mask of the target object's true pixels (only for image datasets) to remove the impact of participants' imprecision or hand jitter that might have included the background pixels. 
The exact segmentation masks for images are created by two authors and included in the benchmark. 
Human attention masks for image datasets are stored in the format of grayscale masks the same size as original images. 
The human attention masks for text datasets are JSON objects with lists of index-word pairs with human-attention scores in the range of 0 to 1.0. 
We did not perform any feature filtering for text annotation samples.
The benchmark is stored in a public domain and free for research use.

\begin{figure*}[t!]
\centering
    \begin{subfigure}[b]{1.0\columnwidth}
        \includegraphics[width=0.8\columnwidth]{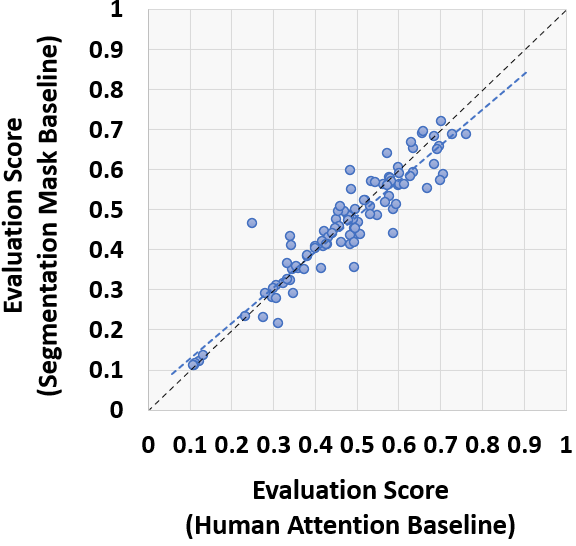}
        \centering \caption{Relation between two ground truth measures}
    \end{subfigure}
    \hfill
    % \hspace{amount}
    \begin{subfigure}[b]{1.0\columnwidth}
        \includegraphics[width=0.77\columnwidth]{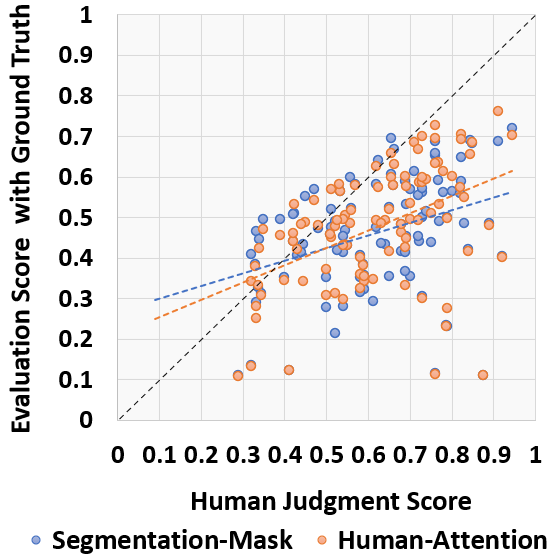}
        \centering \caption{Relation between subjective rating and baseline measures}
    \end{subfigure}
  \caption{Comparison of averaged evaluation scores ($1.0 -$ MAE) between two ground truth baselines and human judgment rating for each sample.
  Evaluation scores are not normalized and the black dashed lines shows the ideal regression line with the slope equal to 1.0 and intercept of zero. 
    (a) Scatterplot of evaluation scores based on segmentation mask (vertical axis) and human-attention mask (horizontal axis).
    (b) Scatterplot of evaluation score based on two ground truth baselines and human judgment rating.}
  \label{fig:results-1}
\end{figure*}

\section{Evaluation Experiments} 

% what 
In this section, we present multiple evaluation experiments to validate the proposed benchmark with empirical results.
These experiments compare three baselines: 1) human-attention mask (our approach) as the ground truth, 2) segmentation mask as the ground truth, and 3) human-judgment rating for evaluating model saliency explanations.
% why 
Our goal is to understand the relationship between the three evaluation methods and communicate the benefits of the proposed benchmark over other common evaluation methods in the literature. 
% how 
The series of experiments are based on saliency maps generated by the Grad-CAM~\cite{selvaraju2017grad} technique for a VGG-19~\cite{simonyan2014very} image classifier on a subset of 100 validation samples from the two classes of cat and dog in PASCAL VOC dataset. 
The VGG network is pre-trained on ImageNet-1k~\footnote{https://pytorch.org/docs/master/torchvision/models.html} and tuned on PASCAL VOC 2007 for the purpose of this evaluation. 
All evaluation scores are based on pixel-wise Mean Absolute Error (MAE) between model saliency score map and the ground truth baseline. 

The saliency map error is calculated as the pixel-wise Mean Absolute Error (MAE) between model saliency score map and the ground truth mask. 
We also looked into False Positive (FP) and False Negative (FN) saliency explanation errors individually.  
We calculate FP saliency error as pixel-wise MAE for the model saliency map scores outside the object's segmentation mask (i.e., error in background pixels) and FN error as the pixel-wise MAE for model saliency map scores inside the ground truth mask (i.e., error in target pixels). 
In the following subsections, we review details and share evaluation results from the three methods.

\subsection{Comparison to Segmentation Mask}

%-what 
In the first evaluation experiment, we compare our proposed human-attention benchmark (multi-layer feature mask) with the segmentation mask (single-layer feature mask) as the evaluation ground truth for the set of saliency maps from Grad-CAM technique.
%-why
Given the lack of granularity for distinguishing important features in the segmentation mask, we hypothesize that the two baselines would result in different evaluation scores for the same set of inputs.

%-how 
Intuitively, the difference between the two baselines is that unlike the segmentation mask, which scores all target features equally, the human-attention mask gradiates the ``salient'' features more than others.
To identify the difference between two evaluation baselines, we calculate evaluation scores using both baselines for direct comparison. 
Specifically, we first normalize both ground truth masks and model saliency maps and then calculate the pixel-wise MAE error between model saliency map and the ground truth baseline. 
For example, a saliency map identical to its human attention mask results in zero MAE error.
In the opposite situation, with cases having no overlap between the ground truth mask and the model saliency map, the MAE error would be 1.0. 
% complete overlap between the ground truth mask and saliency map results in zero MAE error.
Note that MAE is a threshold agnostic metric that---unlike Intersection over Union (IoU)---does not require choosing the $\tau$ hyperparameter for generating objects' binary masks or bounding boxes, see \cite{choe2020evaluating} for more discussion.
Also, evaluating the saliency score map (without converting to a binary mask) retains the granular information in the model explanation.

% \begin{figure}[t!]
% \centering
% \begin{subfigure}[b]{0.70\columnwidth}
%         \includegraphics[width=1.0\columnwidth]{figures/mask-attention-new.PNG}
%         \centering \caption{Relation between two ground truth measures}
%     \end{subfigure}
%       \hfill
%       \\
%     \begin{subfigure}[b]{0.70\columnwidth}
%         \includegraphics[width=1.0\columnwidth]{figures/method-comparison.PNG}
%         \centering \caption{Relation between subjective rating and baseline measures}
%     \end{subfigure}
%   \caption{Comparison of averaged evaluation scores ($1.0 -$ MAE) between two ground truth baselines and human judgment rating for each sample.
%   Evaluation scores are not normalized and the black dashed lines shows the ideal regression line with the slope equal to 1.0 and intercept of zero. 
%     (a) Scatterplot of evaluation scores based on segmentation mask (vertical axis) and human-attention mask (horizontal axis).
%     (b) Scatterplot of evaluation score based on two ground truth baselines and human judgment rating.}
%   \label{fig:results-1}
% \end{figure}

\subsubsection{Results}

% \eric{missing explanations of why...why look at correlation? what does it mean that they're correlated? why do the regression? why compare slopes? what was the rationale and what did you learn from these? don't trust readers to think through it for themselves---walk them through it}

Figure~\ref{fig:results-1}-(a) shows the scatter plot of evaluation scores ($1.0~-$ MAE) between human-attention and segmentation mask baselines. 
The two evaluation scores are statistically significantly ($r = 0.896$ , $p < 0.001$) correlated, as expected. 
Using a linear regression test, we find a regression slope of $w = 0.896$ and intercept of $b = 0. 48$. 
As seen in Figure~\ref{fig:results-1}-(a), this weight and bias result in different evaluation scores between the two ground truth, especially in the higher and lower range of scores. 
To examine the statistical significance of the difference between two ground truth evaluations, we use an ANCOVA test with a custom model to the test for homogeneity of regression slopes between the calculated regression model and the ideal of slope $1.0$ with a zero intercept. 
The test for homogeneity of regression slopes fails with a significant difference ($p < 0.001$) between the two lines indicating that the two evaluation baselines are not equal. 
We also look into FP and FN saliency explanation errors individually.  
The results show that the difference between the two baselines is only due to FN errors being treated differently between the two baselines. 
This was expected since both baselines measure zero evaluation score for the saliency explanations outside the ground truth mask.

\subsection{Comparison to Human Judgment}

%what  
In the second evaluation experiment, we compare explanation evaluation scores using the two ground truth baselines with the human ratings of explanation goodness. 
Subjective human ratings of the model explanations are commonly used as a direct approach for evaluating machine learning explanations by providing a numerical rating of explanations goodness using a simple quantitative measure such Likert scales. 
However, subjective measures typically lack precision and may include user bias.
%why 
We hypothesize that results from human-judgment scores will be significantly different for both (human-attention mask and object segmentation mask) ground truth evaluations. 
% how                                                                   
We use the same subset of images and saliency map explanations from Grad-CAM technique similar to the previous section for the purposes of this human-subjects study. 
Figure~\ref{fig:grad-cam}-(Top) shows examples of heatmap overlays from the Grad-CAM technique used in the user study.

\subsubsection{Human Judgment Interface and Data Collection} 

We designed a simple interface to collect user feedback about the quality of heatmap overlays from the Grad-CAM saliency explanation technique. 
The participants started by reading task instructions followed by a series of images for review and rate. 
% procedure and instruction 
Given an image from the test set, the target classification, and a heatmap overlay, participants were instructed to ``review and rate the heatmaps which explain what parts the AI used to make it's classification decision'' and were asked to rate the ``goodness'' of the AI decision on the scale of 1 to 10. 
A total of 200 unique participants' were recruited from Amazon Mechanical Turk and paid \$0.20 per HIT to review and rate 14 images. 
The first four image ratings (identical images were used for all participants) were used as training and attention check examples; these were disregarded for data collection.

\subsubsection{Results:}

Figure~\ref{fig:results-1}-b shows a scatterplot of the evaluation scores ($1.0~-$ MAE) between human judgment ratings and ground truth scores from objects' segmentation masks and human attention masks. 
The two regression lines for human-attention ground truth (in orange) and segmentation mask (in blue) show both baselines produce different evaluation scores from the user rating scores. 
% The difference indicates that the human subjective rating 
To test for the statistical significance of observed differences, we first normalize user ratings across participants by subtracting each participant's mean rating.  
Then, we use a Pearson's correlation test and linear regression test to compare the human judgment rating scores and the two ground truth scores. 
The user ratings show a moderate-strength correlation with object segmentation mask baseline ($r = -0.121$, $p = 0.002$) and small correlation with human-attention mask baseline ($r = -0.306$, $p < 0.001$).
We also observe signs of user bias, noting that none of the participants rated any of the saliency map instances in the test set below 3-stars even though there are multiple examples with scores below $0.3$ for both ground truth evaluation types. 
These cases were specifically from the examples with multiple occurrences of the target object in which the saliency map was only pointing to one of the target objects. % incomplete explanations
This could potentially indicate a side effect of lower user attention in reviewing cases with incomplete saliency explanations.

To compare measurements between evaluation approaches, we run a linear regression analysis and find that the segmentation mask scores fit with a slope of $w = 0.313$ and intercept of $b =0.268$  (Figure~\ref{fig:results-1}-b, blue trend line), and the fit for human-attention mask scores has a slope of $w = 0.428$ and intercept of $b = 0.210$ (Figure~\ref{fig:results-1}-b, orange trend line). 
Note that the difference between the two linear regression models' slopes with the ideal slope of 1.0 is higher with the segmentation-mask baseline. 
% between evaluation results of the two ground truth methods and human judgment rating,
To examine the statistical significance difference between the measures, we use ANCOVA with a custom model to test for homogeneity of the regression slopes between the two regression models as well as between the calculated regression model and the ideal of slope $1.0$ with zero bias. 
The homogeneity test fails with a significant difference of $p < 0.001$ between the two regression models and the ideal line. 
The analysis indicates the subjective measurement of explanations goodness produces significantly different results from both objective ground truth measures.

\begin{figure}[t!]
\centering
    \begin{subfigure}[b]{0.32\columnwidth}
        \includegraphics[width=0.9\columnwidth]{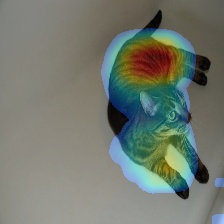}
        \\
    \end{subfigure}
    \begin{subfigure}[b]{0.32\columnwidth}
        \includegraphics[width=0.9\columnwidth]{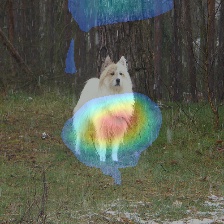}
        \\
    \end{subfigure}
    %   \hfill
    \begin{subfigure}[b]{0.32\columnwidth}
        \includegraphics[width=0.9\columnwidth]{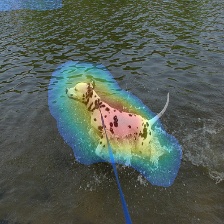}
        \\
    \end{subfigure}
    % \begin{subfigure}[b]{0.24\columnwidth}
    %     \includegraphics[width=0.9\columnwidth]{figures/grad-cam-4.jpg}
    %     \\
    % \end{subfigure}
    \\
    \begin{subfigure}[b]{0.32\columnwidth}
        \includegraphics[width=0.9\columnwidth]{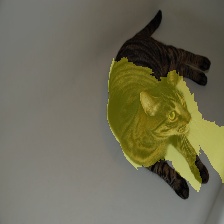}
        \\
    \end{subfigure}
    \begin{subfigure}[b]{0.32\columnwidth}
        \includegraphics[width=0.9\columnwidth]{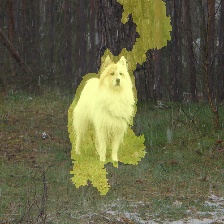}
        \\
    \end{subfigure}
    %   \hfill
    \begin{subfigure}[b]{0.32\columnwidth}
        \includegraphics[width=0.9\columnwidth]{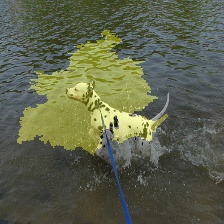}
    %     \\
    % \end{subfigure}
    % \begin{subfigure}[b]{0.24\columnwidth}
    %     \includegraphics[width=0.9\columnwidth]{figures/lime-4.jpg}
        \\
    \end{subfigure}
  \caption{Examples of heat-map overlay of saliency maps using  \textbf{(Top)} Grad-cam and \textbf{(Bottom)} LIME techniques used in the user study for human judgment.}
  \label{fig:grad-cam}
\end{figure}

\begin{figure*}[t!]
\centering
\begin{subfigure}[b]{1.0\columnwidth}
        \includegraphics[width=0.79\columnwidth]{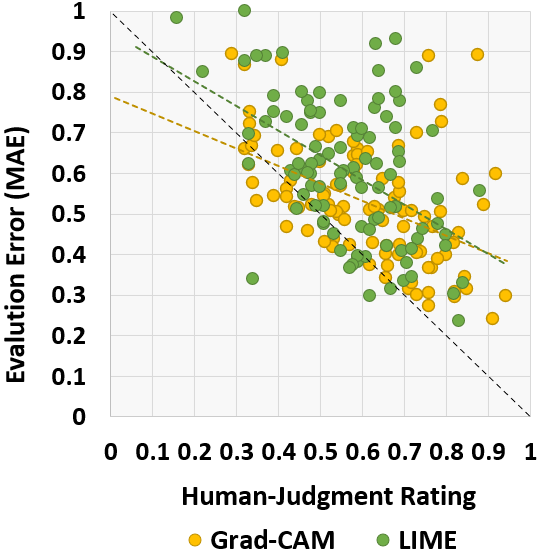}
        \centering \caption{Subjective and ground truth evaluation scores from LIME and Grad-CAM Explanations.}
    \end{subfigure}
    \hfill
    % \hspace{10mm}
    \begin{subfigure}[b]{1.0\columnwidth}
        \includegraphics[width=0.8\columnwidth]{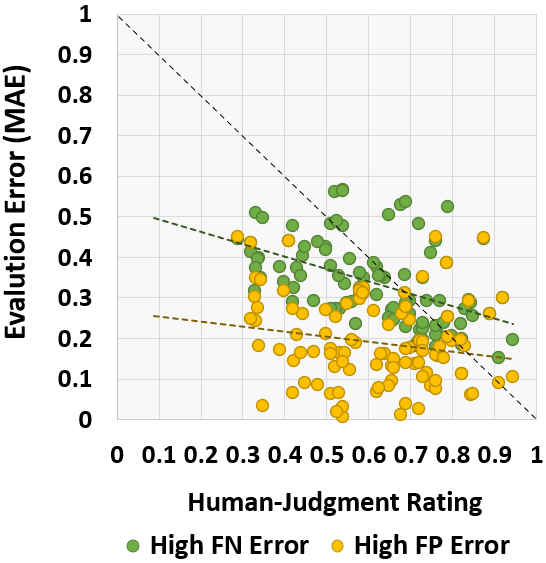}
        \centering \caption{Subjective and ground truth evaluation scores for samples with high FP and High FN explanation MAE error.}
    \end{subfigure}
  \caption{Discrepancies between averaged human judgment rating of saliency explanations and human-attention baseline evaluation. 
  Evaluation scores are not normalized and the black dashed lines shows the ideal regression line with the slope equal to -1.0 and intercept of zero. 
  \textbf{(a)} Participants evaluate saliency explanations from LIME and Grad-CAM differently. 
  \textbf{(b)} Participants evaluate saliency explanations' FP error (model looking at background pixels) differently than FN errors (model not looking at target pixels).}
  \label{fig:results-2}
\end{figure*}

\section{Discussion}

In this section, we review and discuss the evaluation experiments and open problems around model explanation evaluation. 
The evaluation experiment results showed that the human-attention benchmark has allowed for a higher level of granularity in the evaluation of saliency maps and reflected human attention to the features in comparison to the single-layer object's segmentation map.
As compared to the human judgment rating evaluations, we observed signs of participants' bias in their ratings. 
%  Hypothesis: For LIME vs. Gradcam should show same results in VGG-16
%  Hypothesis: For VGG-16 vs. Tuned-VGG-16 show show higher results.

%  Recap 1
\subsubsection{Implications of Results}

% what
We ran human-subject experiments to understand the differences between the subjective and objective evaluation of saliency explanations.  
Although the evaluation results from the three methods had positive correlations, the experimental results showed significant differences among all evaluation scores. 
The difference in scores was mainly due to the clear non-uniform distribution of weights in human attention masks while the segmentation mask weights are uniformly distributed for all features (e.g., pixels, words). 
% Results suggest that both the segmentation mask and the human-attention baselines generate similar scores for FP explanation errors, though there are clear differences.
% Although there are clear differences between human-attention baseline \jeremy{scores?} and segmentation mask baseline \jeremy{scores?}, it's clear that human-attention follows a non-uniform feature attribution while the segmentation mask features are uniformly distributed, since they are derived by the machine.

While segmentation mask benchmarks are mainly used for object segmentation evaluation and weakly supervised object localization~\cite{li2018tell,choe2020evaluating}, the human-attention baseline reflects human factors in feature attribution. 
For example, in annotations of living things, users were more likely to select facial features as important features while the segmentation mask offers a uniformly weighted single-layer mask. 
This is reflected in the evaluation results with human judgment with participants' ratings of explanations being closer to the human-attention baseline rather than the segmentation mask baseline. 
Due to the same effect, evaluation results with the human-attention baseline could be extended to better anticipate user acceptance and trust in model explanations when putting on different applications.

% \begin{figure}[t!]
% \centering
% \begin{subfigure}[b]{0.70\columnwidth}
%         \includegraphics[width=1.0\columnwidth]{figures/lime-grad_cam.PNG}
%         \caption{}
%     \end{subfigure}
%       \hfill
%     \begin{subfigure}[b]{0.70\columnwidth}
%         \includegraphics[width=1.0\columnwidth]{figures/user-bias.PNG}
%         \caption{}
%     \end{subfigure}
%   \caption{Discrepancies between averaged human judgment rating of saliency explanations and human-attention baseline evaluation. 
%   Evaluation scores are not normalized and the black dashed lines shows the ideal regression line with the slope equal to -1.0 and intercept of zero. 
%   \textbf{(a)} Participants evaluate saliency explanations from LIME and Grad-CAM differently. 
%   \textbf{(b)} Participants evaluate saliency explanations' FP error (model looking at background pixels) differently than FN errors (model not looking at target pixels).}
%   \label{fig:results-2}
% \end{figure}

%  Result 1
\subsubsection{Human Biases in Rating}

We explored the human judgment evaluation results to find other possible external or internal factors that could affect participants' subjective ratings. 
For example, human judgment ratings may include user biases toward visual appearance or completeness of saliency maps resulting in incorrect ratings. 
% \TODO{Prior work...} examples include participants' prior-knowledge and expectations that could interfere with their ratings. 
We reviewed and compared the results from human judgment for Grad-CAM and LIME explanations to identify possible biases. 
Also, we reviewed the results to assess possible participants' biases toward model explanation FP and FN error types.

% bias toward visual appearance of saliency maps
To evaluate the effect of visual appearance of saliency explanations, we compare participants' rating of saliency map explanations from LIME~\cite{ribeiro2016should} technique to Grad-CAM explanations on the same subset of images and the same classifier. 
The saliency explanations from the LIME technique (Figure~\ref{fig:grad-cam}-(Bottom)) are visually more chunky and pixelated (mainly due to use of super pixels in LIME's pipeline) compared to smooth concept activation maps from Grad-CAM technique (Figure~\ref{fig:grad-cam}-(Top)).
We analyze results after running a new user study to collect participants' subjective ratings of LIME explanations.

% super pixel defines one ... .

We used two linear regression models to compare participants' ratings between the two groups, see~\ref{fig:results-2}-(a).
We find the slope of $w = -0.428$ and intercept of $b =0.789$ for the user ratings on samples with LIME saliency map (Figure~\ref{fig:results-2}-(a) green trend line) and slope of $w = -0.607$ and intercept of $b = 0.947$ for samples with Grad-CAM saliency map (Figure~\ref{fig:results-2}-(a), yellow trend line). 
We would have expected to see the similar regression slopes between the two groups if the users were evaluating both saliency map explanation types similarly. 
However, the test for homogeneity between the two regression slopes shows a significant difference ($p < 0.001$) between the two model error types.
This indicates that users rated the saliency maps differently, although ground truth evaluation score (Figure~\ref{fig:results-2}-(a), y axis) for both sets of samples. 
% LIME: y = -0.428x +  0.789; 
% Grad-cam: y = -0.607x +  0.947; 

% bias toward saliency error types 
We then analyze participants' rating behavior with respect to different explanation error types. 
We first divided the samples for the test set into two groups with high FP (when the model is looking at background pixels) explanation error and high FN explanation errors (when the model is missing foreground pixels). 
% We then used linear regression models to compare participants' ratings between the two groups. 
Using linear regression models, we find the slope of $w = -0.121$ and intercept of $b =0.265$ for the samples with FP explanation error score (Figure~\ref{fig:results-2}-(b) yellow trend line) and slope of $w = -0.306$ and intercept of $b = 0.525$ for samples with high FN explanations error score (Figure~\ref{fig:results-2}-(b), green trend line). 
We would have expected to see the similar regression slopes between the two groups if the users were evaluating both saliency error types similarly. 
However, the test for homogeneity between the two regression slopes shows a significant difference ($p < 0.001$) between the two explanation error types.
This indicates that users pay less attention to FP explanation errors and in turn, are more critical for FN explanation errors. 
Looking at image samples from the user study, these images included several examples in which the target object was on a smaller scale and the model saliency map was largely exceeded to the background pixels. 

% \TODO{discuss the similarity between user bias in this study and the effect of placebic explanations in previous work}

% Discussion 1
% \subsubsection{Reproducibility and Objectivity Trade-off between Baselines}
\subsubsection{Reproducibility and Objectivity Trade-off}

One way to categorize different evaluation measures is by their objectivity and reproducibility of results. 
As implemented in this paper, users' subjective rating of explanations could collect results for correctness and goodness of model generated explanations.
Ribeiro et al.~\shortcite{ribeiro2016should} presented a case for correction of model explanation in which users reject wrong features and add new features for quantitative evaluation of model explanations.
% the difference (e.g., IoU) between the model explanation and the user-edited explanation could give a precise measure of quality for the model explanation. 
A different method is to collect user feedback through the direct comparison of explanations from multiple interpretability techniques. 
For example, users could review several options to choose the best machine-generated explanation and provide justifications for their choices.  
However, although these methods can provide detailed insights, subjective user feedback is not reusable for new models and interpretability techniques. % generate different explanations, requiring new human review. 
This limitation indeed exists in studies for evaluating XAI systems in different applications and domains~\cite{doshi2017towards}, including tasks and scenarios concerned with the fairness of the decision-making system.

On the other hand, objective evaluation that utilizes ground truth, provides quantitative and reproducible results, yet lacks the guidance of human correctness and goodness scores that show which improvements would be most significant. 
Our benchmarks bridge the trade-off between objectivity and subjectivity of a baseline to satisfy both evaluation aspects.

\subsubsection{Limitations and Future Work}

A limitation of creating this human-attention benchmark is the annotation cost for multi-level human explanation masks.
However, annotation cost could be justified when compared to repeated novel rounds of evaluations for subjective human judgments.
As typically, the iterative design and evaluation of machine learning based systems require multiple rounds of training and test. 
Our human attention benchmark can significantly reduce evaluation costs over design cycles. 
Further, an open question in creating human-attention benchmarks is how to standardize all annotators' perception of explanation when performing the annotation task.
% what aspects of explanation evaluation are beyond the scope of what you're presenting here
% --------------- Future work --------------
In our future work, we plan to study annotators’ behavior on objects in different size and pose to learn general patterns in human attention. 
This could potentially help to optimize the number of annotators for each sample. 
Lastly, we are interested in examining the use case of the human-attention benchmark for tuning models to improve prediction rationale and its effects on explanation quality.

\section{Conclusion}

We present a new model-explanation evaluation benchmark for multiple datasets in image and text domains. 
Our benchmark is designed for quantitative evaluation of saliency map explanations based on human attention on features.
This human-grounded benchmark enables fast, replicable, and objective execution of evaluation experiments for saliency explanations. 
We studied the relationships and trade-offs between two different human-grounded evaluation approaches (i.e., single-layer annotation mask and human subjective feedback) to present the efficiency of the proposed human attention baseline.
Our study results indicated the difference between threshold-agnostic evaluation with a human-attention baseline as compared to previous methods with binary ground truth masks and labels. 
Our experiments also revealed user biases on different explanations' visual appearance and error types in the subjective rating of explanations.

\section{Acknowledgements}
% This research is based on work supported by the \textit{Anonymous Grant.}
This research is based on work supported by the DARPA XAI program under Grant \#N66001-17-2-4031 and NSF award \#1900767.
% DARPA XAI program under Grant \#N66001-17-2-4031. 
% We want thank our Amazon Mechanical Turk ...

\bibliographystyle{aaai}
\bibliography{ref}

\end{document}